# Estimating phylogenetic trees from genome-scale data

Liang Liu[1,2], Zhenxiang Xi[3], Shaoyuan Wu[4], Charles Davis[3], and Scott V. Edwards[4*]

[1]Department of Statistics, University of Georgia, Athens, GA 30602, USA

[2]Institute of Bioinformatics, University of Georgia, Athens, GA 30602, USA

[3]Department of Organismic and Evolutionary Biology, Harvard University, Cambridge, MA 02138, USA

[4]Department of Biochemistry and Molecular Biology & Tianjin Key Laboratory of Medical Epigenetics, School of Basic Medical Sciences, Tianjin Medical University, Tianjin 300070, China.

*Corresponding author:

Scott Edwards

Department of Organismic and Evolutionary Biology

Harvard University

Cambridge, MA 02138, USA

Email: sedwards@fas.harvard.edu

**Abstract**

As researchers collect increasingly large molecular data sets to reconstruct the Tree of Life, the heterogeneity of signals in the genomes of diverse organisms poses challenges for traditional phylogenetic analysis. A class of phylogenetic methods known as 'species tree methods' have been proposed to directly address one important source of gene tree heterogeneity, namely the incomplete lineage sorting or deep coalescence that occurs when evolving lineages radiate rapidly, resulting in a diversity of gene trees from a single underlying species tree. Although such methods are gaining in popularity, they are being adopted with caution in some quarters, in part because perceived shortcomings, poor performance, model violations or philosophical issues. Here we review theory and empirical examples that help clarify these conflicts. Thinking of concatenation as a special case of the more general case provided by the multispecies coalescent model (MSC) can help explain a number of differences in the behavior of the two methods on phylogenomic data sets. Recent work suggests that species tree methods are more robust than concatenation approaches to some of the classic challenges of phylogenetic analysis, including rapidly evolving sites in DNA sequences, base compositional heterogeneity and long branch attraction. We show that approaches such as binning, designed to augment the signal in species tree analyses, can distort the distribution of gene trees and are inconsistent. Computationally efficient species tree methods that incorporate biological realism are a key to phylogenetic analysis of whole genome data.

The emergence of phylogenomic data provides unprecedented opportunities to resolve challenging phylogenies of species and, ultimately, the Tree of Life. In the last few years, a number of phylogenetic and population genetic methods for analyzing the evolutionary history of whole genomes have emerged. These new methods are gradually replacing methods that were once the bulwarks of evolutionary genomics and molecular ecology in the PCR-era. For example, recently the classic isolation-with migration model of phylogeography, originally introduced by Hey, Nielsen and others[1-3] and widely used in likelihood and Bayesian formats, has been updated to accommodate whole genome data[4,5]. Like most methods in phylogeography, these methods model genomic data as a series of unlinked or partially linked loci whose histories are influenced by the underlying demographic history of the species in question. The coalescent provides a robust and general framework for many of these new genome-scale models in phylogeography.

Like models in phylogeography, genome-scale methods in phylogenomics are undergoing a transition, grappling with the heterogeneity of signals that frequently emerge from genome-scale data[6,7]. Recent research has revealed a surprising array of such heterogeneous signals, including variation among loci in base composition, evolutionary rate, and, perhaps most conspicuously, topological congruence[8-12]. Indeed, mirroring early insights into gene tree heterogeneity discovered in the 1980s in the context of phylogeography, gene tree heterogeneity in phylogenetics has emerged as a ubiquitous element, particularly as the number of loci in phylogenetic studies has increased[6]. Yet, surprisingly, whereas phylogeography has dealt with this heterogeneity by acknowledging and modeling stochasticity – indeed, statistical phylogeography has not known any other means of modeling such variation – for decades such models made few in roads into phylogenetics. Even as heterogeneity in gene trees was acknowledged as a significant issue in phylogenetics[13], concatenation or supermatrix methods provided the main paradigm in which phylogenetic models were developed to deal with diverse signals found in multilocus data[14]. Perhaps more intriguing, then, is that, even as such so-called 'species tree' or coalescent models have been developed in the context of phylogenetics, they are being embraced by phylogeneticists cautiously, or in some cases openly questioned[15,16]. Such caution and questioning is no doubt healthy, but it also suggests that, for some researchers, the heterogeneity observed in gene trees is either deemed unimportant or inconsequential for phylogenetic analysis, or that the models developed thus far to deal with this heterogeneity are unsatisfying, incomplete or flawed.

Phylogenomic data have intensified debates over whether concatenation or coalescent methods in phylogenomics are more appropriate for analyzing multilocus sequence data[6,15,17,18]. Although recent phylogenomic studies suggest that the majority of relationships yielded by concatenation and coalescent trees are consistent with each other, or differ from each other without high statistical support[11,12,19,20], recent examples of highly supported conflicting relationships favored by concatenation and coalescent methods have highlighted the details, weaknesses, and assumptions of both sets of methods[11,21-26]. The significant differences in performance of concatenation and coalescent methods in estimating species trees flow directly from the distinct assumptions on which the two methods are based[13,27,28]. Perhaps more importantly, it can be shown that under certain conditions the coalescent model reduces to the concatenation model[18] (see below), which can help explain the frequent similarities between concatenation and coalescent trees in empirical data analyses. Finally, in comparing concatenation and coalescent methods, important questions have been raised regarding the performance of coalescent methods under genetic forces such as within-gene recombination and gene flow, or issues with data, such as sampling effort and strategies, missing data, misrouting of gene trees, binning (concatenating) of multiple loci or errors in gene tree estimation[15,16,29-33]. Whereas population genetic effects such as gene flow and recombination rarely figured in discussions of phylogenetic models in the supermatrix era, recent discussions of these topics appear to be a direct result of the advent of coalescent methods and mark a growing appreciation by phylogeneticists of the links between population genetics and phylogenetics[34,35]. Overall, the current debates offer a glimpse of a field in transition, grappling with new signals and heterogeneity brought on by phylogenomics.

In this review we discuss a number of recent trends in the application of coalescent models to phylogenetic analysis, and address some recent criticisms of such models. Most coalescent models in phylogenomics assume simple models of instantaneous speciation, in which no gene flow occurs after species begin to diverge. Moreover, these models assume complete neutrality, no recombination within loci and free recombination between loci, such that loci can be treated as independent neutral replicates conditional on the phylogenetic history of the lineages under study. Several recent papers have shown that violations of these assumptions can have varying effects on the outcome of phylogenetic analysis (Table 1), with some more severe than others. The main criticisms of coalescent methods in phylogenetics – articulated most forcefully in a series of papers and comments by Gatesy and Springer[15,16,21] – focus on features of phylogenomic data sets that are perceived to violate the multispecies coalescent model (MSC). Such concerns focus on several issues, including the potential for recombination within loci, particularly for transcriptome data in which exons that might span megabases in the genome yet are 'concatenated' together, either *in silico* or by the cell during the process of transcription; the claim that different species tree methods yield conflicting results when applied to the same data sets; the suggestion that most gene tree heterogeneity results from effects other than incomplete lineage sorting (ILS); and confusion over the effects and implications of low-resolution gene trees on species tree estimation. This last issue has also motivated the development of add-on methods to species tree estimation, such as naïve and statistical binning[36,37], whose aim is to augment phylogenetic signal when inferring species trees using coalescent methods.

Here we discuss these various criticisms and concerns in an effort to clarify a host of issues raised about coalescent models in phylogenetics. We also highlight recent empirical studies that evaluate the signal in gene trees and explore their implications for the adoption or rejection of species tree methods. We show that concatenation methods can be considered a special case of the more general MSC. As such, the more specific concatenation model is expected to yield results that are biased but with a smaller variance (e.g., higher bootstrap support) for the estimates of model parameters than the more general MSC -- a perspective that helps explain recent trends in observed phylogenomic data sets. We also suggest that naïve binning – a proposal to augment the signal in multilocus data sets by concatenating genes at random into bins – will only work under highly restricted conditions, and that the lower performance of coalescent methods versus concatenation methods in those studies reflects an undue focus on point estimates and a restricted set of simulation conditions, an undervaluation of the variance of those estimates and the often inflated support of concatenation analyses. Overall we find that currently proposed species tree methods represent a promising start to the challenge of analyzing phylogenomic data and highlight conceptual and practical challenges for the future.

### Concatenation versus Coalescent Models in Phylogenomics

The coalescent and concatenation models differ in their treatment of individual gene trees. In the coalescent model, which assumes free recombination among genes, the gene trees are treated as conditionally independent random variables $G = \{g_i, i = 1, ..., k\}$ given the species tree $S^{38}$, where $k$ is the number of genes. However, independent gene trees may have the same topology, especially when the species tree has long internal branches in coalescent units[39,40]. Under the coalescent model, the likelihood function of the species tree $S$ given the multilocus sequence data $D = (d_1, d_2, ..., d_k)$ is given by

$$L(S|D) = \int_G f(D|G,\lambda) \times \phi(G|S) dG . \quad (1)$$

In (1), $f(D|G,\lambda)$ is the probability density function of sequence data $D$ given gene trees $G$ and parameters $\lambda$ in the substitution model, and $\phi(G|S)$ is the coalescent distribution function of gene trees $G$ given the species tree $S$. The function $f(D|G,\lambda)$ is the traditional likelihood function used for building maximum likelihood gene trees. When gene trees $G$ are identical with the species tree $S$ (i.e., $g_1 = g_2 = ... = g_K = S$), the coalescent probability function $\phi(G|S)$ is 1 for $G = S$ and 0 otherwise. Therefore, when all gene trees are identical with the species tree $S$, the likelihood function of the species tree $S$ is reduced to

$$L(S|D) = f(D|S,\lambda) . \quad (2)$$

As model parameters $\lambda$ may or may not be linked across genes (i.e., partitions), $f(D|S,\lambda)$ in (2) is the likelihood function for the concatenation model with or without partitions. Equation (2) shows that the concatenation model with or without partitions is a special case of the coalescent model. This result has at least two important implications: (i) while concatenation methods may produce inconsistent estimates of species trees under the coalescent model[28], coalescent methods can consistently produce the true species tree under the concatenation model, and (ii) the comparison between the coalescent and concatenation methods falls into the general bias-variance dilemma, i.e., the reduced model (concatenation) in general is biased and has smaller variance for the estimates of model parameters[41].

As a reduced model, concatenation has a smaller number of parameters, because all gene trees in the concatenation model are treated as the same parameter. Thus, the estimates of parameters in the concatenation model tend to have a smaller variance. Since small variance corresponds to high bootstrap support or posterior probability, overestimation of bootstrap support by concatenation methods is a consequence of the fact that it is a reduced model. Moreover, gene-tree-based coalescent methods – those methods that estimate the species tree using separately estimated gene trees[42,43] – can in turn result in larger variance for their estimates of species trees, because these methods estimate gene trees and the species tree separately in two steps using summary statistics or a pseudo-likelihood function. This is the same as saying that gene-tree-based coalescent methods may sometimes result in bootstrap support that is lower than those coalescent methods using the full coalescent model, such as BEST[27] and *BEAST[44]. (Indeed, we have observed this empirically for the 30-locus data set from birds[45] analyzed by Liu and Pearl[27], which yielded a posterior probability of 0.9 when analyzed by BEST but only ~54% bootstrap support when analyzed by the simpler model in STAR[46]). However, the problem of larger variance in gene-tree-based coalescent methods can be greatly alleviated by either improving the efficiency of these methods or by increasing the number of genes. Because phylogenomic data often contain hundreds of genes, species tree analyses using gene-tree-based coalescent methods have produced highly supported species trees for empirical phylogenomic data[11,20,23,24,47]. In contrast, inconsistency of concatenation tends to become severe when there are a large number of genes[28]. Thus, the major concern of phylogenomic data analysis is not the high variance of gene-tree-based coalescent methods, but rather the inconsistency of concatenation methods due to model misspecification[43], especially in cases where phylogenomic inferences are based on highly supported relationships.

In empirical phylogenomic data analyses, concatenation and coalescent methods often produce similar relationships for the majority of branches in the estimated trees. As discussed above, when the concatenation model applies, we do not expect to encounter highly supported relationships that conflict between concatenation and coalescent trees, because with the assumption of identical gene trees, the coalescent model is reduced to the concatenation model. Thus, highly supported but conflicting relationships in concatenation and coalescent trees, as has recently been observed in several phylogenomic data analyses[24,48], indicate either a high amount of ILS (due to short internal branches [in coalescent units] in the species tree) that mislead the concatenation method[11], or that the major cause of gene tree variation is not ILS. In the latter case, neither concatenation nor coalescent models can adequately explain the distribution of heterogeneous gene trees.

**Empirical examples of conflict between coalescent and concatenation methods**

*Coalescence versus concatenation in plant phylogenomics*

To date, very few studies have utilized coalescent methods for inferring plant phylogenies[30,49-51]. Of these studies, strongly conflicting relationships involving concatenation versus coalescent analyses have recently been demonstrated for key nodes in land plant phylogeny. Despite tremendous effort, relationships between the five main seed plant clades–angiosperms, conifers, cycads, *Ginkgo*, and gnetophytes–have remained uncertain. A first broad coalescent analysis of seed plants by Xi et al.[23] incorporated 305 nuclear genes and suggest an explanation for why concatenation methods may result in strong topological incongruence, manifested as phylogenetic 'flip-flops' between analyses involving different subsets of data. Unlike most previous analyses using concatenation, which have strongly placed cycads and *Ginkgo* as successive sisters to the remainder of extant gymnosperms, coalescent results instead strongly identified *Ginkgo* and cycads as monophyletic. Suspecting that the rate of nucleotide substitution might be influencing this difference, Xi et al.[23] binned sites into fast and slow evolving categories and reanalyzed these data. Coalescent analyses continued to support the monophyly of *Ginkgo* and cycads regardless of rate category, but concatenation did not. Instead, fast evolving sites strongly supported the more traditional placement of cycads and *Ginkgo* whereas slow evolving sites supported the placement inferred from coalescent methods. This finding raised the hypothesis that rate variation among sites may explain the striking topological differences observed in concatenated phylogenomic analyses, reflecting problems commonly encountered in, for example, cases of 'long branch attraction'[52,53]. Furthermore, more recent coalescent analyses using expanded taxon sampling and transcriptome data corroborated the monophyly of *Ginkgo* and cycads [50].

In a second paper, Xi et al.[24] further explored this hypothesis by examining a slightly larger phylogenomic dataset including 310 nuclear genes, mostly from flowering plants. Here, phylogenetic relationships were congruent between concatenation and coalescent methods, except for the placement of *Amborella*, which has long been heralded as the sister to all other flowering plants[54,55]. Here, coalescent analyses consistently and strongly support the less traditional placement of *Amborella* as sister to water lilies across all nucleotide rate partitions. By contrast, concatenation showed the same kind of strongly conflicting results that were observed in earlier study involving the placement of *Ginkgo* and cycads: slow evolving sites corroborated the results from coalescent analyses and fast evolving sites placed *Amborella* alone as the first lineage of extant plants. An additional assessment of these fast evolving sites showed particularly strong evidence of saturation, suggesting one explanation for these artifacts. In this case, it appears that distributing saturated sites among many individual gene trees and analyzing each separately as is done by commonly used coalescent methods may be more effective at diluting the deleterious effects of such characters. By contrast, when analyzing all such sites simultaneously within a single matrix as is done using concatenation, such deleterious effects will be exacerbated. We should note that this question has been revaluated more recently with coalescent analyses using expanded taxon sampling and transcriptome data [50]. This study strongly supports the more traditional placement of *Amborella* alone. However, one of us (Z.X.) has analyzed the nearly 400 transcripts common to both *Amborella* and water lilies using these data. Here again, slow evolving nucleotide and amino acid sites strongly support the placement of *Amborella* and water lilies as monophyletic (unpublished data).

Xi et al. investigated this placement of *Amborella* further with simulations. The first of these simulations used the *Amborella* dataset. Here, they randomly constrained each gene tree to be consistent with one of the alternative placements of *Amborella*. Branch lengths and substitution model parameters were then estimated on each constrained tree from the original data. Data sets were then simulated on these trees and the resulting simulated data were binned into fast and slow evolving sites. Coalescent analysis of these data reconstructed the constrained topologies as expected, regardless of rate category. Surprisingly, however, despite 60-80% of the gene trees constrained to the *Amborella* plus water lilies placement, concatenation of the fast evolving sites still resulted in the placement of *Amborella* and water lilies as successive sisters to all angiosperms. This strongly suggests that concatenation of fast evolving sites likely plays a strong role influencing the misleading placement of *Amborella*.

Additional simulations[24] on species trees with both long and short branches suggested that when incomplete lineage sorting is high, concatenation methods perform very poorly, suggesting that concatenation may strongly be influenced by the shape of the species tree topology and its interaction with ILS. A related follow-up simulation study [25] suggests that when long external and short internal branches occur simultaneously with high ILS, concatenation methods can be misled, especially when two of these long branches are sister lineages. By contrast, species tree methods (in this case MP-EST[43] and STAR[46]) are more robust under these circumstances. This result is particularly relevant because many ancient radiations across the Tree of Life suggest this particular pattern of adjacent long and short branches. Because short internal branches in the species tree can increase the potential for ILS and gene tree discordance, these results indicate that coalescent methods are more likely to infer the correct species tree in cases of rapid, ancient radiations where short internal and long external branches are in close phylogenetic proximity.

*Coalescence versus concatenation in mammal phylogenomics*

Despite recent progress in classifying eutherian mammals into four superorders – Afrotheria, Xenarthra, Laurasiatheria and Euarchontoglires, several key relationships within eutherian mammals remain controversial, including the root of Eutheria, and the interordinal relationships within Euarchontoglires, Laurasiatheria and Afrotheria[56,57]. To date, however, the reconstruction of mammalian phylogeny has relied mostly on concatenation methods, which as we indicate above may suffer from systematic bias due to the unrealistic assumption of gene tree homogeneity across loci. To empirically address the effect of gene tree heterogeneity on estimating deeply diverging phylogenies, Song et al.[48] took the approach of subsampling loci and taxa so as to investigate the robustness of concatenation and coalescent methods to different analyses of the same taxa. Using a data set of 447 nuclear genes for 35 mammalian taxa, Song et al. demonstrated that concatenation indeed behaves inconsistently across data sets, as evidenced by the conflicting and strongly supported relationships from different subsamples of loci. In contrast, coalescent methods were able to estimate a consistent phylogeny for eutherian mammals from the same subsets of data, and demonstrated clear positive relationship between nodal support values and the number of loci. In this regard, the study on mammals was consistent with predictions of a recent simulation study that showed a correlation between number of loci and species tree support for nodes exhibiting high rates of ILS[31].

Several studies [58], including the mammal study[11], have revealed the sometimes striking contrast between support values for trees analyzed by concatenation versus coalescent methods, with the latter often yielding much lower values even when the tree is largely similar. It is often the case that a tree that appears well-resolved by concatenation methods is found to be poorly resolved using coalescent methods. So which set of support values better reflects reality? Such results likely reflects the tendency for concatenation methods, especially Bayesian concatenation methods, to overestimate credibility values in phylogenetic trees[59,60]. They may also partly reflect the tendency of gene-tree-based coalescent methods to yield lower confidence levels than full Bayesian coalescent methods when the number of gene trees is small (see above). Differences in the method of bootstrapping may also contribute to these discrepancies; the multilocus bootstrap[61] is generally believed to more accurately capture support in large data sets than the simple bootstrap[16]. In the study by Song et al.[11], the 26 genes that resolved a sample of mammalian taxa using concatenation resulted in a poorly resolved tree when analyzed by MP-EST. They suggested that, in a coalescent framework, approximately 400 genes would be required to resolve the sample of taxa in the particular tree for mammals. In this case, missing data may also play a role, since coalescent methods appear to be more sensitive to missing data than concatenation methods[58]. However, one study[62] suggested that species tree methods were "remarkably resilient to the effects missing data". In our view, what sensitivity to missing data displayed by species tree methods reflects the true impact of missing data on phylogenetic analysis -- an impact that is obscured by analyses employing concatenation. There are now many examples of well resolved phylogenies employing concatenation on highly incomplete data sets[50,63,64]. We believe these examples illustrate the power of concatenation to obscure the true support for trees by the collected data (see also[65]).

## Maximizing signal in species tree analysis

Species tree analysis has moved past the stage of uncritical adoption to evaluation of sampling strategies and methods for maximizing phylogenetic signal[29,34,66-69]. Much of our understanding of the behavior of species tree reconstruction comes from simulation studies, although analyses of empirical data have also yielded important insights. An important rule of thumb that has emerged from both simulation and empirical studies is that species trees are only as good as the gene trees on which they are built[10,34,58,66,70,71]. This maxim applies both to 'two-step' species tree methods, in which gene trees are used as input data, as well as to 'single step' approaches, such as Bayesian methods, in which gene and species trees are estimated simultaneously. For example, several empirical studies on organisms as varied as turtles, mammals, fish and flowering plants, have shown that species tree estimation can be misled by biased gene tree estimation due to long-branch attraction and base compositional heterogeneity among lineages, a manifestation of substitution model non-stationarity[8,10,24]. Even so, recent work suggests that species tree methods, even those in which gene trees are estimated first and separately from the species tree, may be less susceptible to classic challenges in phylogenetic analysis, such as long branch attraction (Table1)[25]. This lowered susceptibility of species tree methods may be due to the fact that a typical gene tree is based on at most a few thousand base pairs, which may be small enough such that departures from stationarity may be less visible in the underlying data sets. While it is valid to criticize the ensemble of constituent approaches that comprise gene tree and species tree analysis, including gene tree and species tree reconstruction, it is unwarranted to criticize species tree methods per se, especially when it is the reconstruction of gene trees that is responsible for misestimation[16]. Thus a number of authors have suggested improving and maximizing signal in gene tree estimation as a means of improving species tree estimation as a whole. Such improvements in signal take a variety of forms, including binning of subsets of genes, using longer or more informative genomic regions for each locus [11,26,71,72], minimizing base compositional heterogeneity among lineages, and even choosing genes with specific trends in base composition, for example those genes trending towards AT richness in mammals[73]. In this section we evaluate a variety of suggestions for signal enhancement in species tree analysis.

*Naïve and statistical binning*

The naïve binning technique was proposed to improve the support of coalescent estimates of species trees by reducing the estimation error of gene trees[37]. This technique concatenates DNA sequences across randomly selected genes, regardless of whether the selected genes share the same history. The binned sequences are treated as a "super gene", and used to estimate "super gene trees". As the binned sequences are longer than the original data, the resulted gene trees are often well supported[37]. However, as shown in Kubatko and Degnan[28], binning sequences from genes with distinct histories can mislead maximum likelihood (ML) concatenation methods, which consistently produce the wrong estimate of the species tree under broad conditions. This inconsistency problem may also occur for the "super genes" in the binning technique. Indeed, although the paper title implied that binning yielded a general improvement for phylogenomic data ("Naive binning improves phylogenomic analyses"), in the paper the authors were more equivocal ("This paper should not be interpreted as recommending the use of naïve binning, but instead as an indication of the potential for binning techniques to improve species tree estimation"; p. 2284). From first principles we know that, for the case of 4-taxon anomalous species tree (see Figure 3f in Kubatko and Degnan[28]) when the bin size (BS) is large, all of the concatenated genes (i.e., super genes) will have the same ML tree, a tree that is incongruent with the species tree. These biased super gene trees can significantly mislead the MP-EST estimate of the species tree. Even when the bin size is small (10-15 genes), binning sequences can increase the probability of estimating gene trees that disagree with the species tree. When this probability is greater than a threshold, it will mislead the MP-EST method to consistently produce the wrong estimate of the species tree.

To demonstrate these phenomena, we simulated gene trees from a 5-taxon species tree (labeled species A–E, Figure 1a) under the coalescent model using the function *sim.coaltree.sp* in the R package Phybase[74]. Because the MP-EST method assumes that gene trees are rooted trees, species E is used as the outgroup for rooting the estimated gene trees. To reduce the rooting error, we intentionally set a small population size $\theta = 0.01$ and a long internal branch (length = 0.08) between the ingroup species (A-D) and the outgroup species E. Under these conditions, the ingroup species (A-D) almost always form a monophyletic group in the simulated gene trees. In addition, the population size parameter $\theta$ is set to 0.1 for other ancestral populations in the species tree. This species tree is in the anomaly zone, because the internal branches for species A, B, C, and D are very short (0.005/0.1 = 0.05 in coalescent unit), and the most probable gene tree (PT) does not match the species tree[75]. DNA sequences of length 1000 bp were generated from the simulated gene trees using SeqGen[76] with the Jukes-Cantor (JC69) model[77]. The simulated DNA sequences were binned at random to form super genes. The bin size was set to 10, 20, 30, 40, and 50 genes, respectively. A ML super gene tree was estimated for each bin by PhyML[78]. The estimated super gene trees were then used to estimate species trees using the MP-EST method. Each simulation was repeated 100 times.

When the bin size is one (i.e., no binning was performed), the probabilities of the estimated gene trees are similar to the true probabilities of gene trees generated from the species tree (Figure 1b). However, the probability of the most probable gene tree (PT) significantly increases as the bin size increases (Figure 1b). When the bin size is 30, almost half of the estimated super gene trees are PT, which is incongruent with the species tree (Figure 1b). It is clear from this exercise that binning sequences from genes with distinct histories can bias the distribution of the estimated gene trees, with a high probability of producing the tree that is incongruent with the species tree.

Without binning (BS = 1), the MP-EST method can consistently estimate the correct species tree as the number of genes increases (Figure 2a). The proportion of trials yielding the correct species tree appears to increase as the number of genes increases, and reach 1.0 when the number of genes is 1000. When the bin size is five, the probabilities of estimating the correct species tree are greater than those without binning. This result indicates that when the bin size is five, binning can improve the performance of MP-EST in estimating species trees. However, when the bin size is greater than or equal to 10, the probability of estimating the correct species tree is in general less than the probability without binning (Figure 2b). Moreover, the probability of estimating the correct species tree appears to decrease as the number of bins (i.e., the number of super genes) increases (Figure 2b). For bin sizes of 30, 40, and 50, the probability of yielding the correct species tree decreases to zero when the number of bins reaches 80 (Figure 2b). Meanwhile, when the bin size is greater than 10, the MP-EST tree based on the binned genes appears to consistently estimate the wrong tree as the number of bins increases (Figure 2c). Interestingly, for the simulation parameters studied here, for a bin size of 10, the probability of the correct tree stabilizes around 0.6, while the probability of an incorrect tree stabilizes around 0.4. Although this result suggests that when the bin size is less than 10, the binning technique can be beneficial for improving the performance of the MP-EST method in estimating species tree, additional simulation results suggest that this may not always be the case (not shown). Further research on naïve binning is needed. We can certainly state that when the bin size is greater than 10, binning sequences across genes with distinct histories can significantly bias the distribution of estimated gene trees, and result in inconsistent estimates of species trees.

Ideally, the loci should be concatenated if no or only a few recombination events occurred between those loci. A model based on biology would suggest that binning should be based on loci that are closely linked in genomes, such as often occurs in transcriptomes, because the chance of recombination is positively related to the physical distance between two loci. Recently, Mirarab et al[36] proposed a statistical binning technique which attempts to bin loci with the same gene tree. In this approach, loci are binned when there are no strongly supported topological conflicts among the estimated gene trees of those loci, for example when the gene trees do not conflict on branches with > 75% bootstrap support. However, the statistical properties of statistical binning are not yet fully explored. As discussed in the previous section, the assumption of free recombination between genes and no recombination within genes plays a key role in the coalescent model. Loci are treated as conditionally independent, due to the assumption of free recombination between genes. Two loci may have the same gene tree even though they are conditionally independent. This may occur, for example, when the species tree has long internal branches (in coalescent units). When the genes have the same history, binning their sequences can improve gene tree estimation, but it also reduces the sample size of independent genes. Thus, from first principles we can state that binning sequences from genes with the same history may not necessarily improve species tree estimation. In addition, two trees with no strongly supported conflicts do not necessarily indicate that they are topologically identical with each other. As we discussed for the naïve binning, combining loci with different histories may seriously bias the distribution of the estimated gene trees, which in turn will mislead species tree estimation. In conclusion, further studies are needed to evaluate the performance of statistical binning.

*Information content of genes for species tree reconstruction*

Despite the recent arrival of genome-scale Bayesian phylogenetic methods for concatenation [79], the Bayesian or approximate Bayesian coalescent model cannot presently be applied to phylogenomic data due to excessive computational burden[27,44,80]. Alternative gene-tree approaches, including MP-EST[43], GLASS[81], Maximum Tree[43], STAR[46], STEM[82], STELLS[83], ASTRAL[84] and STEAC[46], build species trees from estimated gene trees. These approaches have computational advantages that allow them to be used for phylogenomic data analyses[70] and novel molecular markers such as ultraconserved elements[19,72]. Additionally, recent species tree methods utilizing information from SNPs[85,86] and haplotypes[87] may be scalable to large data sets. However, the tree estimates given by gene-tree-based approaches often suffer the problem of big variance (i.e., low bootstrap support). As these approaches employ bootstrap techniques to account for errors in estimating gene trees, large estimation error for gene trees can greatly lower the bootstrap support of the species trees estimated by those approaches (Huang et al 2010).

Moreover, these approaches estimate species trees based only on the topologies of gene trees, ignoring the branch length information, which further reduces the efficiency of those approaches. The simulation study by Huang et al (2010) suggests that high amount of gene tree estimation error may be the major cause of low bootstrap support in estimates of species trees using STEM (Table 1). It has been suggested by empirical studies that using highly supported gene trees (average bootstrap support value > 0.5) can improve bootstrap support of species trees estimated by gene-tree-based approaches[11,17], and adding poorly supported gene trees (i.e., weak genes) does not contribute more information regarding the phylogeny of species. We evaluated this hypothesis through a simple simulation analysis.

To evaluate the effect of weak genes on the performance of gene-tree-based approaches in estimating species trees, DNA sequences were simulated from the true species tree ((((A:0.002, B:0.02):0.002, (C:0.002, D:0.002):0.002):0.002, E:0.006):0.01, F:0.016) with $\theta$ = 0.008. The population size parameter $\theta$ is constant across branches of the species tree. Specifically, gene trees were generated from the species tree under the MSC, again using Phybase[74]. DNA sequences were generated from the simulated gene trees under the JC69 model using SeqGen[76] with the Jukes-Cantor model[77]. We generated 1000 base pairs for 'strong' genes, and 100 base pairs for 'weak' genes. The average bootstrap support values of strong genes range primarily from 70% (first quantile) to 91.75% (3$^{rd}$ quantile) with median = 81.83% (Figure 3a), whereas the average bootstrap support values for 100 base pair genes are mostly < 50% (Figure 3a). Thus we selected 100 base pair genes with bootstrap values < 50% as weak genes. Species trees were reconstructed from 10, 20, up to 90 (in increments of 10) estimated gene trees for strong genes using MP-EST. Each simulation was repeated 100 times. The proportion of trials estimating the true species tree was 0.33 for 30 strong genes, and it increased to 1 when the number of strong genes was 40 (Figure 3b). However, when adding weak genes to the set of 30 strong genes, the proportion of estimating the true species tree increased slowly to 0.63 and then decreased to 0.50 (Figure 3b). This result suggests that adding weak genes may not contribute more information to an otherwise strong data set when estimating species trees. When adding weak genes, the distribution of estimated gene trees becomes flat, and the coalescent signal contained in the estimated gene trees is significantly reduced. Thus adding weak genes may actually reduce the performance of species tree estimation methods, negating the old adage that "more data is always better".

*Transcriptomes, base composition, and location-aware concatenation*

Transcriptomes have formed an important type of data lending itself to species tree estimation. Because of their ease of alignment and characterization, transcriptomes will continue to be an important data type for many kinds of phylogenetic analysis. However, because transcriptomes consist solely of coding regions, they are potential targets of natural selection[88]. Indeed, genome-wide phylogenetic comparisons in primates have shown that the rate of ILS in coding regions is lower than that in non-coding regions[88]. This higher incidence of reciprocal monophyly and lower incidence of ILS in coding regions is likely driven by recurrent bouts of positive selection on coding regions, with surrounding noncoding regions adhering to patterns more consistent with neutrality. Whether or not this type of departure from the neutral coalescent will be problematic for species tree analysis is unclear, because by lowering the rate of deep coalescence, natural selection could help eliminate some of the discordance that is known to decrease phylogenetic signal in species tree analysis[42,89]. However, recent genome-wide analyses of avian phylogeny suggest substantial convergence in protein coding regions, resulting in species trees that are clearly incongruent with noncoding genomic partitions [12].

Transcriptomes are also characterized by strong base compositional biases, and it is well known that the third positions of nuclear genes can evolve rapidly and become misleading over long time scales. Departures from base compositional stationarity in phylogenomic data sets have long been known to cause serious problems for phylogenetic analysis, potentially linking together lineages that are unrelated but share similar base compositions[12,90-92]. Variation in base composition among taxa is known for both transcriptome data and for whole-genome data. Additionally, it is well known that different subgenomes can possess different base compositions; for example, coding regions are generally more GC-rich than noncoding regions. Examination of variation in base composition among lineages and genes as well as saturation patterns in different data partitions has shed some light on optimal choice of marker for species tree analyses. For example, Chiari et al.[10] showed that species tree analysis of turtle relationships changed dramatically depending on whether amino acid sequence or nucleotide sequence data were used to build gene trees, with amino acid data sets providing more congruent results. Betancur-R et al.[8] showed convincingly that substantial apparent gene tree heterogeneity in fish data sets arises from mis-estimation of gene trees most likely due to base compositional heterogeneity among lineages. They found that choosing sets of genes with base compositional homogeneity among lineages substantially reduced the apparent evidence for gene tree heterogeneity and ILS. While we view this result as significant, we suggest that it does not invalidate the use of species tree methods or necessarily support supermatrix approaches, because species tree methods do not require gene tree heterogeneity to work well. A final example in which base composition plays a role in species tree reconstruction was provided by Romiguier et al.[73], who found that the gene trees in mammals based on GC-rich coding regions were more heterogeneous than those with an AT-bias. Moreover, the species tree suggested by AT-rich data sets favored the Afrotheria hypothesis, with Afrotheria as the first branch within placental mammals, whereas those based on the more heterogeneous GC-rich data sets favored the Atlantogenata rooting, in which Xenartha and Afrotheria are sister groups. These authors suggested that the way forward was not so much to account for ILS using novel phylogenetic methods as to reduce its incidence by using particular sets of markers. Again, while we do not dispute the observation of lower gene tree heterogeneity in AT-rich markers, this does not mean that supermatrix approaches are necessarily favored. So long as different genes are conditionally independent from one another due to recombination, we suggest that species tree analysis will still accumulate phylogenomic signal very differently, and we hypothesize more accurately, than will concatenation approaches. Species tree approaches are not invalidated by the absence of ILS; rather they represent a fundamental recognition of the importance of stochasticity from gene to gene that is unaccounted for by concatenation methods, even when ILS is low or absent.

An important issue in the ongoing discussion of concatenation versus coalescent methods is whether the location-aware concatenation of exons in transcriptome data is reasonable or whether it violates the MSC[21]. Location-aware concatenation occurs when adjacent exons or genomic regions are concatenated to one another. This type of concatenation has biological realism in so far as adjacent regions of the genome are known to be correlated in their historical ancestry, with stretches of chromosomes yielding information suggesting similarity in gene trees as one moves along the chromosome. On the one hand, transcriptome data has been shown in multiple studies to yield phylogenetic trees produced by MSC models that are either congruent with previous results or provide novel hypotheses that are plausible (e.g., turtles, plants). On the other hand, transcriptome data is indeed 'concatenated' by cells when converting pre-mRNAs into the mature mRNAs that are often used in phylogenomic studies. A key difference between the location-aware concatenation performed by cells and the naïve binning recommended by Bayzid and Warnow[37] is that, in location-aware concatenation, exons that are adjacent to one another in the genome are concatenated, whereas binning approaches are agnostic as to the location of binned genes in the genome. Still, exons that are concatenated by cells can still occur at varying distances from one another in the genome, and may experience levels of recombination in their history that rivals those experienced by loci located at a distance in the genome. However, as pointed out by Lanier and Knowles[35] point out, recombination will only be a challenge on extremely short internal branches of the species tree; recombination occurring on long branches will involve sequences closely related within species and will provide discordant signal primarily when taking place in common ancestral species. The empirical effects of varying genomic distance on the variety of observed gene trees needs to be studied in more detail. Using a variety of statistical methods, several studies, particularly in primates and rodents, have found a patchwork of gene tree signals in chunks when moving along a chromosome[93-95]. Methods for delimiting genomic segments that display consistency of phylogenetic signal are emerging, and these may prove extremely useful for delimiting loci for species tree analysis and minimizing the negative effects of recombination. In general, however, when faced with insufficient signal in a species tree analysis, we advocate increasing the size of phylogenomic data sets[72], rather than pseudoconcatenation or naïve binning as a means of augmenting signal. Despite the plethora of large-scale phylogenomic studies, empiricists have not yet exhausted phylogenetic information, and until then, data collection, rather than binning uninformed by genomic context, should be the method of choice for data augmentation.

*Computational trade-offs in species tree analysis*

In addition to concerns about marker choice and the statistical properties of species tree estimation methods, another major concern is computational cost. When phylogenomic data include thousands of genes, the total length of the concatenated sequences will begin to explode, resulting in an extremely high computation burden when using concatenation methods. For such data sets, it is practically impossible to perform bootstrap concatenation analyses (with just 100 replicates), or model selection analysis for choosing the best substitution model for the data. For some cases, it is even challenging to perform a single maximum likelihood analysis for the concatenated sequences. Recent analyses of genome-scale data have been unable to complete computation for concatenated date sets, and phylogenetic inferences are often made in the absence of analyses that have reached convergence or have searched tree space substantively (e.g., [12,96]). Recent advances in computer architecture of phylogenetic analyses may help alleviate the challenges of analyzing genome-scale supermatrices[97]. On the other hand, Bayesian coalescent models[27,44] have the same computational issues as concatenated analyses when the sequence data involve hundreds of genes. Simpler coalescent methods such as MP-EST, STAR, and STEAC rely on estimated gene trees to infer species trees, and their computational costs are manageable, even for thousands of genes and species. Of course, as a price for computational efficiency, gene-tree-based methods suffer low statistical efficiency, in the sense that they often require more loci to produce a highly resolved tree. In practice, it is extremely difficult to estimate the sample size for phylogenetic methods such that they can achieve reasonable bootstrap support for their estimates of the species trees. Clearly the landscape of phylogenetic methods for genome-scale data, whether for supermatrices or unlinked loci, is rapidly changing.

**Conclusion**

Phylogenetic analysis of genome-scale data inevitably invites the use of methods that acknowledge the stochasticity of gene histories, and the MSC provides a robust framework for incorporating the information in this stochasticity. Yet newer methods for analyzing genome-scale data, relying on supertrees[98] or alignment graphs[99], will likely yield further insights. The justification for species tree methods lies not in the ubiquity of gene tree heterogeneity in empirical data sets, although this heterogeneity has certainly spurred the advent of such methods. Rather, the justification lies in their acknowledgement of fundamental genetic processes inherent in all organisms, including recombination along the chromosome, which renders gene histories independent of one another, conditional on the phylogeny, and genetic drift, which generates stochasticity in gene tree topologies and branch lengths. Thus, even when all gene trees are topologically similar, species tree methods will yield results differing from concatenation methods, if not in phylogenetic topology then often in phylogenetic support, because species tree methods better model the accumulation of signal that is accrued with increasingly large data sets. Different methods of species tree inference incorporate different amounts of detail of the multispecies coalescent process, and there is a trade-off between model accuracy and computational burden. For now, 'two-stage' species tree methods, in which estimated gene trees are used as input data, are useful in so far as they can analyze large-scale genome-wide data sets with ease. But more complex and computationally efficient models are sorely needed[7]. Although concatenation and species tree approaches often yield similar estimates of phylogeny, an increasing number of examples of strong conflict between concatenation and coalescent analyses shows that the conditions for conflict among methods occur in empirical data. Additionally, concatenation methods appear more sensitive to classic phylogenetic challenges such as long-branch attraction and rate variation among lineages than are gene-by-gene species tree analyses. Some of the differences in behavior between concatenation and species tree methods can be understood as manifestations of the classic bias-variance problem in statistics, because concatenation is a special case of the more general model used by species tree methods, and therefore can exhibit low variance (such as high tree support) despite being more biased than species tree approaches. Further studies aimed at understanding the connections between concatenation and species tree methods and the types of data that maximize signal under the multispecies coalescent model will allow phylogenetics to take full advantage of the flood of data in the genomics era.

**Acknowledgements**

This work was supported by the United States National Science Foundation [DMS-1222745 to L.L. and DEB-1120243 to C.C.D.].

40.     Degnan, J. H. & N. A. Rosenberg. 2006. Discordance of species trees with their most likely gene trees. Public Library of Science Genetics. 2: 762-768.
41.     Kelchner, S. A. & M. A. Thomas. 2007. Model use in phylogenetics: nine key questions. Trends in Ecology & Evolution. 22: 87-94.
42.     Liu, L.*, et al.* 2009. Estimating species phylogenies using coalescence times among sequences. Syst Biol. 58: 468-477.
43.     Liu, L., L. Yu & S. Edwards. 2010. A maximum pseudo-likelihood approach for estimating species trees under the coalescent model. BMC Evol Biol. 10: 302.
44.     Heled, J. & A. J. Drummond. 2010. Bayesian inference of species trees from multilocus data. Mol Biol Evol. 27: 570-580.
45.     Jennings, W. B. & S. V. Edwards. 2005. Speciational history of Australian grass finches (*Poephila*) inferred from 30 gene trees. Evolution. 59: 2033-2047.
46.     Liu, L.*, et al.* 2009. Estimating species phylogenies using coalescence times among sequences. Syst Biol. 58: 468-477.
47.     Faircloth, B. C.*, et al.* 2012. Ultraconserved elements anchor thousands of genetic markers spanning multiple evolutionary timescales Syst Biol. 61: 717-726.
48.     Song, S.*, et al.* 2012. Resolving conflict in eutherian mammal phylogeny using phylogenomics and the multispecies coalescent model. Proc. Natl. Acad. Sci. USA. 109: 14942-14947.
49.     Zhao, L.*, et al.* 2013. Phylogenomic analyses of nuclear genes reveal the evolutionary relationships within the BEP clade and the evidence of positive selection in Poaceae. Plos One. 8: e64642.
50.     Wickett, N. J.*, et al.* 2014. Phylotranscriptomic analysis of the origin and early diversification of land plants. Proc Natl Acad Sci U S A. 111: E4859-4868.
51.     Weitemier, K.*, et al.* 2014. Hyb-Seq: Combining target enrichment and genome skimming for plant phylogenomics. Appl Plant Sci. 2: 1400042.
52.     Felsenstein, J. 2003. Inferring Phylogenies. Sinauer Associates, Inc. Sunderland, MA.
53.     Felsenstein, J. 1978. Cases in which parsimony or compatibility methods will be positively misleading. Systematic Zoology. 27: 401-410.
54.     Amborella Genome Project. 2013. The *Amborella* genome and the evolution of flowering plants. Science. 342: 1241089.
55.     Soltis, D. E. & P. S. Soltis. 2004. Amborella not a "basal angiosperm"? Not so fast. American journal of botany. 91: 997-1001.
56.     Murphy, W. J.*, et al.* 2001. Resolution of the early placental mammal radiation using Bayesian phylogenetics. Science. 294: 2348-2351.
57.     Murphy, W. J.*, et al.* 2007. Using genomic data to unravel the root of the placental mammal phylogeny. Genome Res. 17: 413-421.
58.     Thomson, R. C.*, et al.* 2008. Developing markers for multilocus phylogenetics in non-model organisms: a test case with turtles. Mol Phylogenet Evol. 49: 514-525.
59.     Suzuki, Y., G. V. Glazko & M. Nei. 2002. Overcredibility of molecular phylogenies obtained by Bayesian phylogenetics. Proc Natl Acad Sci USA. 99: 16138-16143.
60.     Misawa, K. & M. Nei. 2003. Reanalysis of Murphy et al.'s data gives various mammalian phylogenies and suggests overcredibility of Bayesian trees. J Mol Evol. 57: S290-S296.
30

Table 1. Studies evaluating the robustness of species tree phylogenetic methods to various genetic forces and sampling schemes.

| Topic | Reference number | Conclusions/comments |
|---|---|---|
| General violation of multispecies coalescent model | 100 | Claims the majority of multilocus sequence datasets are a poor fit to the MSC model, although much of the violation stems from fit of substitution model or unknown sources on a minority of genes. |
| Gene flow | 68,101 | • The coalescent method is robust to low levels of gene flow<br>• Concatenation performs poorly relative to the coalescent methods in the presence of gene flow.<br>• Gene flow can lead to overestimation of population sizes and underestimation of species divergence times in species trees. |
| Sampling/mutation | 29,69,89 | • Increased sampling of individuals per species can significantly improve the estimation of shallow species trees.<br>• Sampling more individuals does not significantly improve accuracy in estimating deep species trees. Adding more loci can improve the estimation of deep relationships.<br>• Mutational variance is a major source of error in estimates of species trees. |
| Recombination | 35,102 | • Recombination has minor effect on species tree estimation except on extremely short species trees.<br>• The negative effect sof recombination can be easily overcome by increased sampling of alleles |
| Missing data | 15,58,62 | • Missing data can decrease the support of species tree estimates<br>• Missing data can significantly affect the accuracy of species tree estimation<br>• Species tree methods are "remarkably resilient" to missing data[62] |
| Taxon sampling | 11 | Compared to concatenation, coalescent methods are more robust to poor taxa sampling |
| Long-branch attraction | 25 | Species tree methods more resilient to the effects of long-branch attraction than concatenation methods |
| Random rooting of gene trees | 32,33 | Misrooting of gene trees can mimic the coalescent process |
| Other | 103 | AGTs themselves are unlikely to pose a significant danger to empirical phylogenetic study |

**Figure legends**

Figure 1: Inconsistency of the binning technique. a) the species tree used for simulating gene trees. In this species tree, $((((A:0.01, B:0.01):0.005, C:0.015):0.005, D:0.02):0.08, E:0.1)$, species E is used as the outgroup. To reduce the rooting error, we set a small population size $\theta = 0.01$ and a long internal branch (length = 0.08) between ingroup species A, B, C, and D and the outgroup species E. The population size parameter $\theta = 0.1$ for other ancestral populations in the species tree. b) The probabilities of two estimated gene trees for binned genes. We consider only two trees, the matching tree (MT) and the most probable tree (PT). The bars at 0 represent the true (coalescent) probabilities of MT and PT generated from the species tree under the coalescent model.

Figure 2: the probability of estimating the wrong and correct species trees without binning. A) In these simulation parameters, bin sizes of 0 (no binning) or 5 converge on the correct tree. B) For bin size (BS) 10, the probability of estimating the correct tree is stable around 0.6 as the number of bins increases. C) In contrast, the probabilities of estimating the wrong species tree for BS = 20, 30, 40, 50 increase to 1 as the number of bins increases.

Figure 3: The effect of non-informative genes on the performance of gene-tree-based approaches in estimating species trees. DNA sequence data were simulated from the true species tree. The sequences of 1000 base pairs were generated for strong genes, while 100 base pairs were generated for weak genes. a) the species tree used in the simulation. b) the boxplot of the average bootstrap values for weak and strong genes. c) the effect of different numbers of strong

and weak genes on the performance of MP-EST in estimating species trees. See text for further explanation.

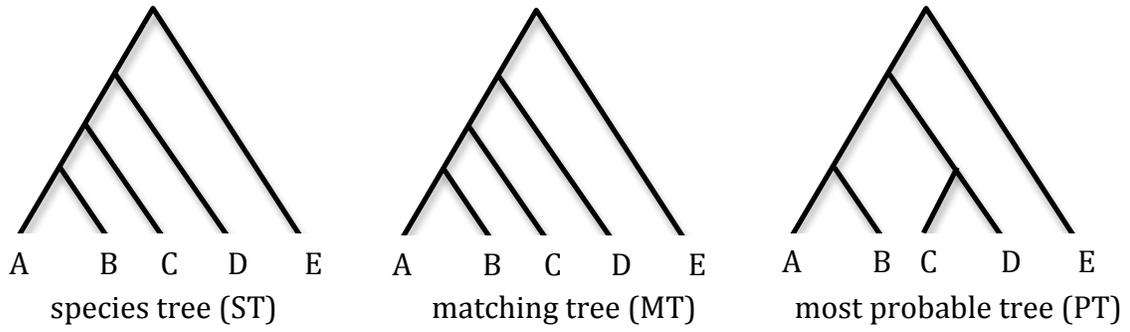

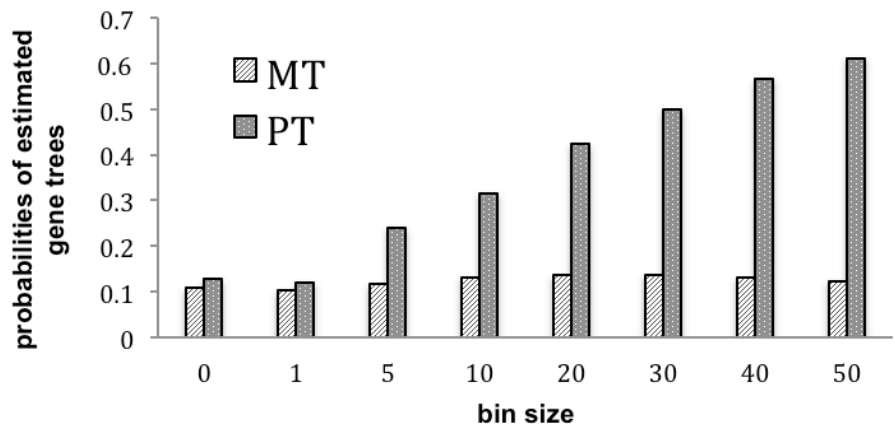

Figure 1

a)

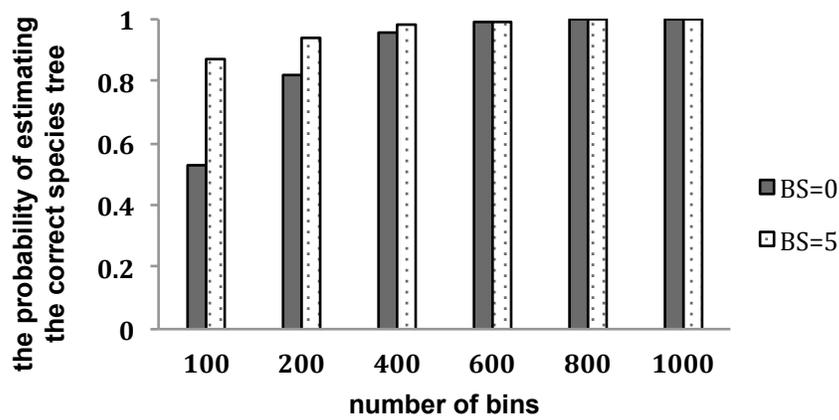

b)

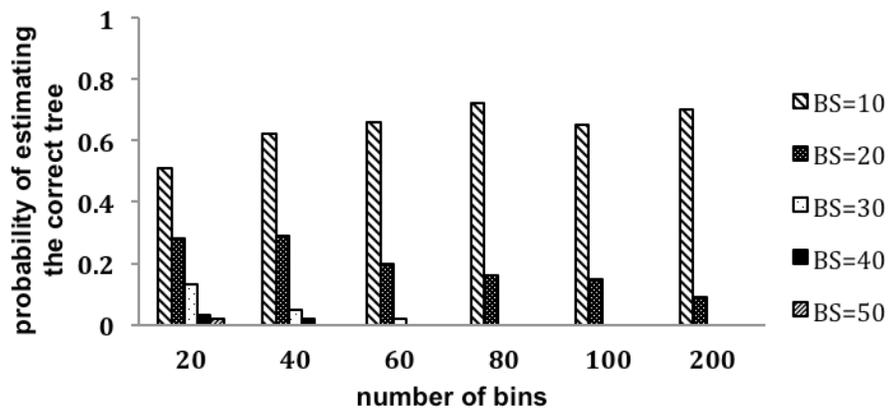

c)

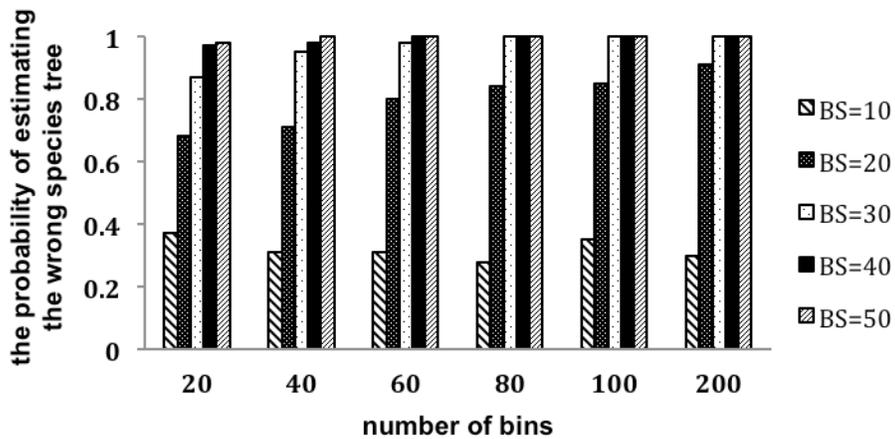

Figure 2

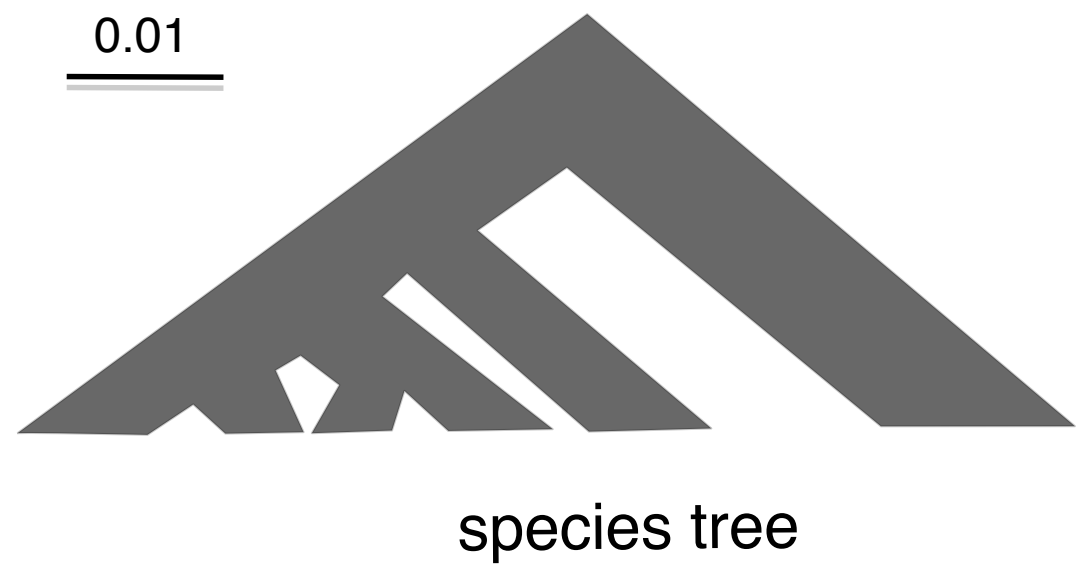 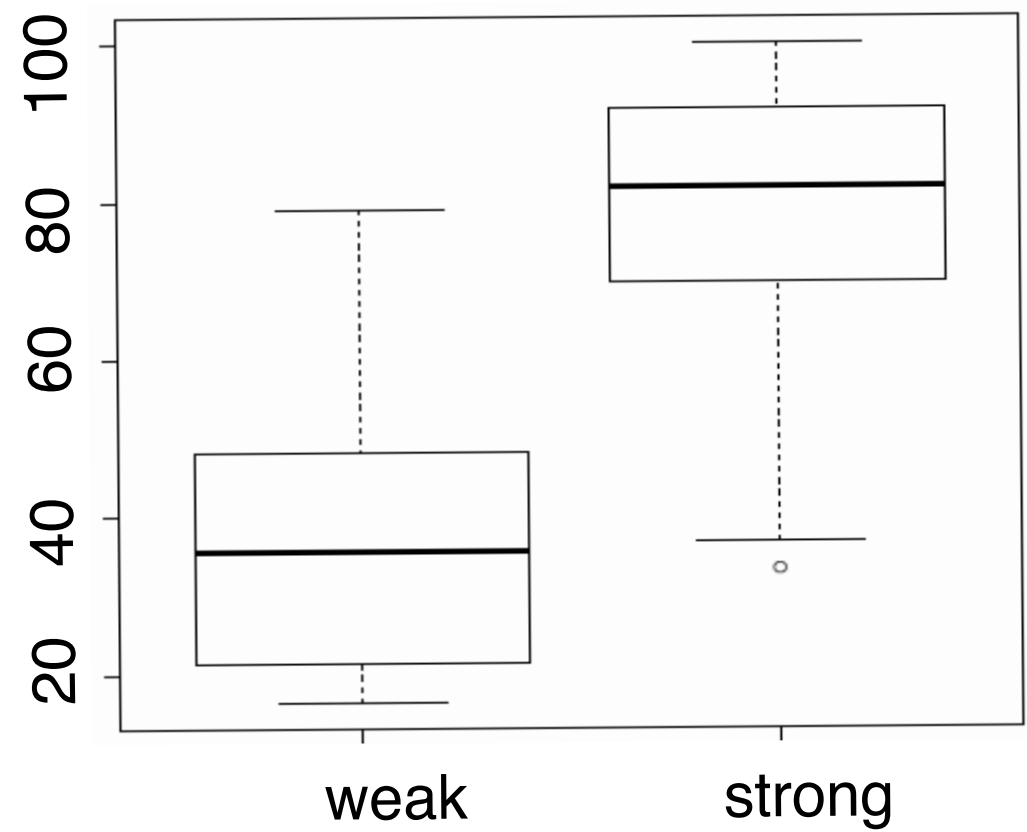 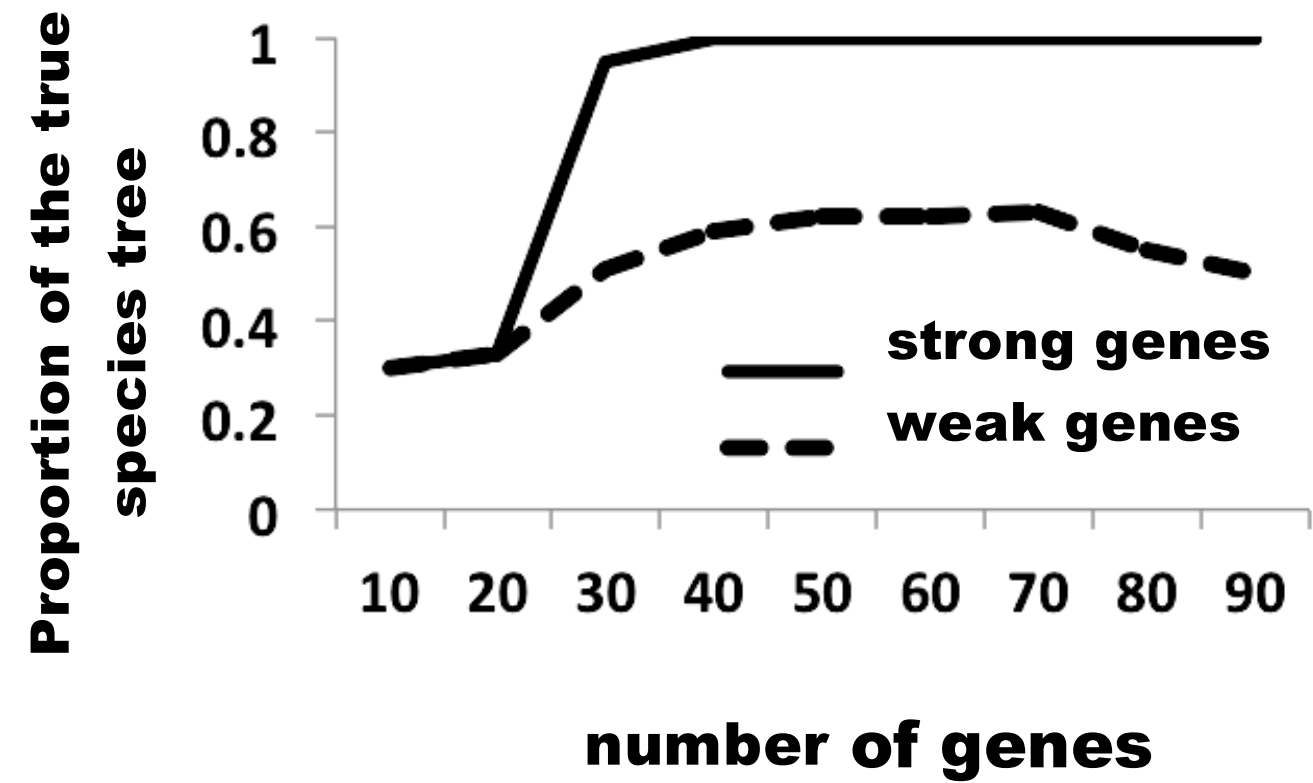

Figure 3